\documentclass[conference]{IEEEtran}

\normalsize

\usepackage[T1]{fontenc}% optional T1 font encoding
\usepackage{amsfonts}
\usepackage{amsmath}
\usepackage{amsthm}
\usepackage{amssymb}
\usepackage{caption}
\usepackage[labelformat=simple]{subcaption}
\usepackage{cite,graphicx,algorithm}
\usepackage{algorithmic}
\usepackage{color}
\usepackage{bm}
\usepackage{makecell}

\ifCLASSINFOpdf
\else
\fi

\theoremstyle{plain}

\newtheorem{lemma}{Lemma}

\newtheorem{remark}{Remark}

\begin{document}
\title{Low-Complexity Memory AMP Detector for High-Mobility MIMO-OTFS SCMA Systems\vspace{-0.36em}}
\author{\IEEEauthorblockN{Yao~Ge$^{1}$, Lei~Liu$^{2}$, Shunqi~Huang$^{2}$, David~Gonz\'{a}lez~G.$^{3}$, Yong~Liang~Guan$^{1}$, and Zhi~Ding$^{4}$}
\IEEEauthorblockA{$^{1}$Continental-NTU Corporate Lab, Nanyang Technological University, Singapore}
\IEEEauthorblockA{$^{2}$School of Information Science, Japan Institute of Science and Technology (JAIST), Nomi 923-1292, Japan}
\IEEEauthorblockA{$^{3}$Wireless Communications Technologies Group, Continental AG, Germany}
\IEEEauthorblockA{$^{4}$Department of Electrical and Computer Engineering, University of California at Davis, Davis, CA 95616 USA}
\IEEEauthorblockA{Emails: $^{1}$\{yao.ge;eylguan\}@ntu.edu.sg, $^{2}$\{leiliu;shunqi.huang\}@jaist.ac.jp, $^{3}$david.gonzalez.g@ieee.org, $^{4}$zding@ucdavis.edu}
}%
% <-this % stops a space
%\thanks{TCOM version based on Michael Shell's bare{\textunderscore}jrnl.tex version 1.3.}}


\markboth{}%
{}



\maketitle

\begin{abstract}
%\boldmath
Efficient signal detectors are rather important yet challenging to achieve satisfactory performance for large-scale communication systems. This paper considers a non-orthogonal sparse code multiple access (SCMA) configuration for multiple-input multiple-output (MIMO) systems with recently proposed orthogonal time frequency space (OTFS) modulation. We develop a novel low-complexity yet effective customized Memory approximate message passing (AMP) algorithm for channel equalization and multi-user detection. Specifically, the proposed Memory AMP detector enjoys the sparsity of the channel matrix and only applies matrix-vector multiplications in each iteration for low-complexity. To alleviate the performance degradation caused by positive reinforcement problem in the iterative process, all the preceding messages are utilized to guarantee the orthogonality principle in Memory AMP detector. Simulation results are finally provided to illustrate the superiority of our Memory AMP detector over the existing solutions.
\end{abstract}


% \fi
\IEEEpeerreviewmaketitle

\section{Introduction}
% \IEEEPARstart{T}{his} demo file

The orthogonal time frequency space (OTFS) modulation \cite{hadani2017orthogonal} has emerged as a promising and alternative modulation scheme to traditional orthogonal frequency-division multiplexing (OFDM) for high mobility communications. Unlike OFDM, OTFS multiplexes information symbols in the delay-Doppler domain and exploits the diversity coming from both the channel delays and Doppler shifts for better performance.
%It can convert a fast time-varying channel in time-frequency domain into a quasi-stationary sparse channel in delay-Doppler domain. 
Thanks to the sparsity of the channel in delay-Doppler domain, the required pilot and receiver complexity for channel estimation can be significantly reduced \cite{li2022joint}. However, the equivalent transmission of OTFS in the delay-Doppler domain involves a sophisticated two-dimensional periodic convolution, leading to severe inter-symbol interference (ISI) \cite{ge2021receiver}. Therefore, simple and effective signal detectors are rather important for OTFS systems to maintain the sufficient diversity of the wireless channel and achieve desired receiver performance.

In addition, multiple-input multiple-output (MIMO) can be combined with OTFS to further increase spectral efficiency and transmission reliability for high mobility scenarios \cite{singh2022low,qu2022efficient,thaj2021low}. Meanwhile, non-orthogonal multiple access (NOMA) has been integrated with OTFS to improve spectrum utilization and massive user connectivity \cite{ge2021otfs,ge2023otfs}. 
%Meanwhile, efficient multiple access protocols represent another strong arena of growth \cite{li2020new,ge2021otfs,deka2021otfs}. A novel path division multiple access (PDMA) is proposed in \cite{li2020new} by scheduling the users in non-overlap angle-delay-Doppler domain for massive MIMO-OTFS networks. To further support massive mobile access, non-orthogonal multiple access (NOMA) has been integrated with OTFS to improve spectrum utilization and user connectivity \cite{ge2021otfs,deka2021otfs}. 
In OTFS-NOMA, multiple users are allowed to access the same delay-Doppler resources simultaneously, and distinguished by different power levels \cite{ge2021otfs} or via sparse code multiple access (SCMA) \cite{ge2023otfs}. 
%In particular, sparse code multiple access (SCMA) \cite{deka2021otfs} has been regarded as a promising technology because of its excellent performance and practical applicability. 
However, a major challenge for MIMO-OTFS multi-user systems is the data detection at the receiver due to the substantial increase of system dimension and the additional inter-user interference, which requires complicated processing with high complexity. 
%Thus, how to efficiently and robustly detect the received MIMO-OTFS multi-user signal with the presence of multi-dimensional interference (i.e., ISI, IDI and IUI), is another challenge to be addressed.

The optimal maximum {\em a posteriori} (MAP) detector is impractical due to its complexity growing exponentially with the system dimension. The existing linear detectors such as zero-forcing (ZF) and linear minimum mean square error (LMMSE) \cite{singh2022low} leading to severe performance loss. 
%However, such linear detectors can not achieve desired performance and still introduce large computational complexity due to the matrix inverse. 
Recently, several studies \cite{9866655,qu2022efficient,thaj2021low} exploit the typical structure and characteristic of OTFS transmission matrix for interference cancellation and performance improvement. Unfortunately, the typical OTFS channel matrix structures do not always hold and the interference cancellation may cause error propagation. By leveraging the sparsity of OTFS channel matrix, the efficient Gaussian message passing (GMP) \cite{raviteja2018interference,xiang2021gaussian} and expectation propagation (EP) \cite{ge2023otfs,9827946} detectors are developed with low-complexity. However, the GMP and EP detectors may suffer from performance loss if so many short girths (i.e., girth-4) exist in the factor graph. 
%Although the damping scheme can relax the effect of short girths and improve the performance, there is no efficient way to obtain the optimal damping solution currently.
Alternatively, orthogonal/vector approximate message passing (OAMP/VAMP) \cite{ge2021otfs} are proposed for OTFS systems with more robustness performance. However, the high computational cost caused by either matrix inverse or singular-value decomposition (SVD) of OAMP/VAMP limits their applications to large-scale systems. 
%To achieve the robust data transmission in large-scale MIMO-OTFS multi-user systems, low-complexity efficient channel equalization and data detection algorithms are desired. 

Recently, a Bayesian inference technique denoted as Memory approximate message passing (AMP) \cite{9805776} was proposed for sparse signal recovery with performance improvement. Inspired by \cite{9805776}, we develop an efficient customized Memory AMP detector for MIMO-OTFS SCMA systems, where the typical sparse channel matrix is applied to further reduce the complexity. Our Memory AMP detector only involves a low-complexity matched filter in each iteration, and applies finite terms of matrix Taylor series to approximate the matrix inverse. 
%We approximate the messages passed between the factor node and variable nodes on the factor graph as Gaussian and further reduce the complexity by taking advantage of the channel matrix sparsity. 
%Based on the specific orthogonality principle, our proposed Memory AMP detector demonstrates the performance and complexity benefits over conventional detector algorithms.
Based on the specific orthogonality principle and closed-form damping solution, our proposed Memory AMP detector outperforms the heuristic damping-based GMP and EP detectors, and achieves similar performance to that of OAMP/VAMP but with low-complexity.

%all the preceding messages are used to ensure that the estimation errors in Memory AMP are asymptotically independent identically distributed (IID) Gaussian. We demonstrate the performance and complexity benefits of the proposed Memory AMP over conventional detector schemes.

\section{System Model}\label{II_model}
%\begin{figure}
%  \centering
%  \includegraphics[width=6.5in]{fig1_diagram.pdf}
%  \caption{Block diagram of the proposed MIMO-OTFS SCMA system.}
%  \label{fig1_diagram}
%\end{figure}
We consider an uplink MIMO-OTFS SCMA system with $J$ independent mobile users transmitting signals to the base station (BS) simultaneously. Without loss of generality, each user is equipped with one transmit antenna and the BS is equipped with $U$ receive antennas. At each transmit slot, every ${\log _2}Q$ information bits ${{\bf{b}}_j}$ from the $j$-th user are mapped into a $K$-dimensional sparse codeword ${{\bf{c}}_j} = {\left[ {c_j^1,c_j^2, \cdots ,c_j^K} \right]^T}$ selected from a user-specific SCMA codebook $\mathbb{A}_j$ of size $Q$, where $j = \{ 1,2, \cdots ,J\} $ and $J>K$ typically, leading to an overloading factor $\zeta  = \frac{J}{K} > 1$. We assume that each user employs only one SCMA layer and only $D$ ($D<K$) non-zero entries among the $K$-dimensional codeword ${{\bf{c}}_j}$. We then generate the information symbols in the delay-Doppler plane ${\bf{X}}_j \in {\mathbb{C}^{M \times N}}$ of the $j$-th user by allocating $\frac{{MN}}{K}$ SCMA codewords ${{\bf{c}}_j}$ either along the delay axis or along the Doppler axis without overlapping.
Here, $M$ and $N$ are numbers of resource grids along the delay and Doppler dimensions, respectively.\footnote{For simplicity, we assume that $M$ and $N$ are integer multiples of $K$, i.e., ${[M]_K} = {[N]_K} = 0$, where ${\left[  \cdot  \right]_k}$ denotes mod-$k$ operation.} The transmission scheme for each user is based on OTFS to combat the doubly-selective fading channels caused by channel delays and Doppler shifts. 
%Finally, the advanced multi-user detection algorithm is employed at the BS to recover the signal of each user.

Specifically, the transmitted time domain signal in OTFS can be obtained by first applying the inverse symplectic finite Fourier transform (ISFFT) on ${\bf{X}}_j \in {\mathbb{C}^{M \times N}}$ followed by Heisenberg transform. Assuming rectangular transmit pulse, the output of the Heisenberg transform can be given by 
\begin{align}
{\bf{S}}_j = {\bf{F}}_M^H({{\bf{F}}_M}{\bf{X}}_j{\bf{F}}_N^H)={\bf{X}}_j{\bf{F}}_N^H,
\end{align}
where ${{\bf{F}}_M} \in {\mathbb{C}^{M \times M}}$ and ${{\bf{F}}_N} \in {\mathbb{C}^{N \times N}}$ represent normalized $M$-point and $N$-point fast Fourier transform (FFT) matrices, respectively. The transmitted time domain signal ${\bf{s}}_j \in {\mathbb{C}^{MN \times 1}}$ can be generated by column-wise vectorization of ${\bf{S}}_j$.

We then add a cyclic prefix (CP) in front of the generated time domain signal for each user. After passing through a transmit filter, the resulted time domain signal of each user is sent out simultaneously over the doubly-selective fading channels. Here, we characterize the impulse response channel between $j$-th user and $u$-th receive antenna as
\begin{align}\label{channel_multipath}
{h_{uj}}\left[ {c,p} \right]
= &\sum\limits_{i = 1}^{L_{uj}} {{h_{uj,i}}{e^{j2\pi {\nu _{uj,i}}\left( {c{T_s} - p{T_s}} \right)}}{{\mathop{\rm P}\nolimits} _\text{{rc}}}(p{T_s} - {t_{j}}-{\tau _{uj,i}})},\nonumber\\& c = 0, \cdots, MN - 1;\; p = 0, \cdots ,P_{uj} - 1,
\end{align}
where ${L_{uj}}$ is the number of multipaths between the $j$-th user and the $u$-th receive antenna, and ${t_{j}}$ denotes the timing offset experienced by the $j$-th user; $h_{uj,i}$, $\tau _{uj,i}$ and $\nu _{uj,i}$ represent the channel gain, delay and Doppler frequency shift associated with the $i$-th path, respectively. The system sampling interval ${T_s} = {1 \mathord{\left/
 {\vphantom {1 {M\Delta f}}} \right.
 \kern-\nulldelimiterspace} {M\Delta f}}$. The Doppler frequency shift $\nu _{uj,i}$ can be further expressed as $\nu _{uj,i} = ({{k_{{uj,i}}} + {\beta _{{uj,i}}}})/NT$, where integer ${{k_{{uj,i}}}}$ and real
${\beta _{{uj,i}}} \in \left( { -0.5,0.5} \right]$ respectively stand for the index and fractional part of $\nu _{uj,i}$. $T$ (seconds) and $\Delta f = {1 \mathord{\left/
 {\vphantom {1 T}} \right.
 \kern-\nulldelimiterspace} T}$ (Hz) are chosen to be larger than the maximal channel delay spread and maximum Doppler frequency shift, respectively.

In (\ref{channel_multipath}), ${{{\mathop{\rm P}\nolimits} _\text{{rc}}}(\cdot  )}$ is an equivalent overall raised-cosine (RC) rolloff filter when the typical root raised-cosine (RRC) pulse shaping filters are applied at the transmitter and receiver. The maximal channel tap $P_{uj}$ is determined by the duration of the overall filter response and the maximum channel delay spread. To overcome the inter-frame interference, we append a CP which is sufficiently long to accommodate both the maximum timing offset and maximal channel delay spread of all users.

The received time domain signal first enters a received filter. After discarding the CP, we can express the received signal from the $j$-th user at the $u$-th receive antenna as
\begin{align*}
r_{uj}[c] = \sum\limits_{p = 0}^{P_{uj} - 1} {h_{uj}[c,p]s_j\left[ {{{\left[ {c - p} \right]}_{MN}}} \right]},\; c = 0, \cdots ,MN - 1.
\end{align*}
The resulting signal ${\bf{r}}_{uj} \in {\mathbb{C}^{MN \times 1}}$ is then devectorized into a matrix ${\bf{R}}_{uj} \in {\mathbb{C}^{M \times N}}$, followed by Wigner transform as well as the symplectic finite Fourier transform (SFFT), yielding 
\begin{align}
{\bf{Y}}_{uj} = {\bf{F}}_M^H({{\bf{F}}_M}{\bf{R}}_{uj}){\bf{F}}_N={\bf{R}}_{uj}{\bf{F}}_N.
\end{align}

Finally, the end-to-end delay-Doppler domain input-output model from $j$-th user to $u$-th receive antenna is given by \cite{ge2021receiver}
\begin{align}\label{relationship_DD}
&Y_{uj}[\ell,k] =  \sum\limits_{p = 0}^{P_{uj} - 1} {\sum\limits_{i = 1}^{L_{uj}} {\sum\limits_{q = 0}^{N - 1} {{h_{uj,i}}{{\mathop{\rm P}\nolimits} _\text{{rc}}}(p{T_s} - {t_{j}}-{\tau _{uj,i}})} } } \nonumber\\
\times &\gamma (k,\ell,p,q,{k_{{uj,i}}},{\beta _{{uj,i}}})X_j[ {{{[ {\ell - p} ]}_M},{{[ {k - {k_{{uj,i}}} + q} ]}_N}} ],
\end{align}
where
\begin{subequations}
\begin{equation}
\begin{aligned}
&\gamma (k,\ell,p,q,{k_{{uj,i}}},{\beta _{{uj,i}}}) 
\\&=\!
\begin{cases}
\frac{1}{N}\xi (\ell,p,{k_{{uj,i}}},{\beta _{{uj,i}}})\theta (q,{\beta _{{uj,i}}}),\!\!&p \!\le\! \ell\! < \!M,\\
\frac{1}{N}\xi (\ell,p,{k_{{uj,i}}},{\beta _{{uj,i}}})\theta (q,{\beta _{{uj,i}}})\phi (k,q,{k_{{uj,i}}}), \!&0 \!\le \!\ell \!<\! p,
\end{cases}
\end{aligned}
\end{equation}
\begin{align}
\xi (\ell,p,{k_{{uj,i}}},{\beta _{{uj,i}}}) = {e^{j2\pi \left( {\frac{{\ell - p}}{M}} \right)\left( {\frac{{{k_{{uj,i}}} + {\beta _{{uj,i}}}}}{N}} \right)}},
\end{align}
\begin{align}
\theta (q,{\beta _{{uj,i}}}) = \frac{{{e^{ - j2\pi ( - q - {\beta _{{uj,i}}})}} - 1}}{{{e^{ - j\frac{{2\pi }}{N}( - q - {\beta _{{uj,i}}})}} - 1}},
\end{align}
\begin{align}
\phi (k,q,{k_{{uj,i}}}) = {e^{ - j2\pi \frac{{{{\left[ {k - {k_{{uj,i}}} + q} \right]}_N}}}{N}}}.
\end{align}
\end{subequations}

The input-output model in (\ref{relationship_DD}) can be further expressed in vector form as
$
{{\bf{y}}_{uj}} = {{\bf{H}}_{uj}}{{{\bf{\tilde x}}}_j},
$
where ${{{\bf{\tilde x}}}_j},{{\bf{y}}_{uj}}\in {\mathbb{C}^{MN \times 1}}$, and ${{\bf{H}}_{uj}}\in {\mathbb{C}^{MN \times MN}}$ is a sparse matrix.
Therefore, the received signal at the $u$-th receive antenna is given by
\begin{align}\label{relation_singlesum}
{{{\bf{\bar y}}}_u} = \sum\limits_{j = 1}^J { {{\bf{H}}_{uj}}{{{\bf{\tilde x}}}_j}}  + {{\bm{\omega }}_u}
=  {{\bf{\bar H}}_u}{{\bf{\bar x}}} + {\bm{\omega }_u},
\end{align}
where $u = \{ 1,2, \cdots ,U\} $, ${{{\bf{\bar H}}}_u} = \left[ { {{\bf{H}}_{u1}}, {{\bf{H}}_{u2}}, \cdots , {{\bf{H}}_{uJ}}} \right]\in {\mathbb{C}^{MN \times MNJ}}$, and ${\bf{\bar x}} = {\left[ {{\bf{\tilde x}}_1^T,{\bf{\tilde x}}_2^T, \cdots ,{\bf{\tilde x}}_J^T} \right]^T}\in {\mathbb{C}^{MNJ \times 1}}$. ${{\bm{\omega }}_u}\in {\mathbb{C}^{MN \times 1}}\sim \mathcal{CN}\left( {{\bf{0}},{\sigma ^2}{\bf{I}}} \right)$ is the complex additive white Gaussian noise (AWGN) at the $u$-th receive antenna. 

Note that ${{\bf{\bar x}}}$ is a sparse vector and the number of non-zero entries in ${{\bf{\bar x}}}$ is only $\frac{{MNJD}}{K}$. Let ${{\bf{x}}}\in {\mathbb{C}^{\frac{{MNJD}}{K} \times 1}}$ denotes the effective input after removing zeros and grouping every $D$ non-zeros elements from the same SCMA codeword in ${{\bf{\bar x}}}$. We also apply similar operations for the corresponding columns in ${{{\bf{\bar H}}}_u}$ to obtain the effective matrix ${{{\bf{H}}}_u}\in {\mathbb{C}^{MN \times \frac{{MNJD}}{K}}}$.

Thus, we can rewrite (\ref{relation_singlesum}) as 
\begin{align}\label{relation_singlefinal}
{{{\bf{\bar y}}}_u} = {{{\bf{H}}}_u}{{\bf{x}}} + {\bm{\omega }_u},\;u = 1,2, \cdots ,U,
\end{align}
where ${\bf{x}} = {\left[ {{\bf{x}}_1^T,{\bf{x}}_2^T, \cdots ,{\bf{x}}_{{{MNJ} \mathord{\left/
 {\vphantom {{MNJ} K}} \right.
 \kern-\nulldelimiterspace} K}}^T} \right]^T}\in {\mathbb{C}^{\frac{{MNJD}}{K} \times 1}}$, ${{\bf{x}}_c}\in {\mathbb{C}^{D \times 1}}$, ${\bf{h}}_{d,c}^u\in {\mathbb{C}^{1 \times D}}$ and $${{\bf{H}}_u} = \left[ {\begin{array}{*{20}{c}}
{{\bf{h}}_{1,1}^u}&{{\bf{h}}_{1,2}^u}& \cdots &{{\bf{h}}_{1,{{MNJ} \mathord{\left/
 {\vphantom {{MNJ} K}} \right.
 \kern-\nulldelimiterspace} K}}^u}\\
{{\bf{h}}_{2,1}^u}&{{\bf{h}}_{2,2}^u}& \cdots &{{\bf{h}}_{2,{{MNJ} \mathord{\left/
 {\vphantom {{MNJ} K}} \right.
 \kern-\nulldelimiterspace} K}}^u}\\
 \vdots & \vdots & \ddots & \vdots \\
{{\bf{h}}_{MN,1}^u}&{{\bf{h}}_{MN,2}^u}& \cdots &{{\bf{h}}_{MN,{{MNJ} \mathord{\left/
 {\vphantom {{MNJ} K}} \right.
 \kern-\nulldelimiterspace} K}}^u}
\end{array}} \right].$$
By stacking the received vectors in (\ref{relation_singlefinal}) as ${\bf{y}} = {\left[ {{\bf{\bar y}}_1^T,{\bf{\bar y}}_2^T, \cdots ,{\bf{\bar y}}_U^T} \right]^T}\in {\mathbb{C}^{UMN \times 1}}$, the input-output model of the MIMO-OTFS SCMA system is given by
\begin{align}\label{relation_central}
{\bf{y}} = {\bf{Hx}} + \bm{\omega },
\end{align}
where ${\bf{H}} = {\left[ {{\bf{H}}_1^T,{\bf{H}}_2^T,\cdots ,{\bf{H}}_U^T} \right]^T}\in {\mathbb{C}^{UMN \times \frac{{MNJD}}{K}}}$ and $\bm{\omega }={\left[ {{\bm{\omega }}_1^T,{\bm{\omega }}_2^T,\cdots ,{\bm{\omega }}_U^T} \right]^T}\in {\mathbb{C}^{UMN \times 1}}$.
%The basic block diagram of the considered system is illustrated in Fig. \ref{fig1_diagram}. 
For convenience, we define $\mathcal{N}=\frac{{MNJD}}{K}$ and $\mathcal{M}=UMN$, respectively.

%From (\ref{relation_central}), one can apply linear receivers such as ZF and LMMSE \cite{9698103} or more advanced non-linear OAMP/VAMP \cite{ge2021otfs} receiver with a large matrix inverse. One can also derive low-complexity GMP \cite{raviteja2018interference,xiang2021gaussian} and EP \cite{ge2021otfs,9827946} receivers with time-consuming empirical damping for multi-user detection. To balance the computational efficiency and receiver performance, we introduce the Memory AMP detector in the next section. 

\section{Memory AMP Detector}\label{III_receiver}

%We now investigate the efficient low-complexity Memory AMP detector to recover the signal of each user from the received signal at the BS. 

\subsection{Memory AMP Detector}

Here, we can use a factor graph to represent the system model of (\ref{relation_central}), where a factor node ${\bf{y}}$ is connected to multiple variable nodes ${{\bf{x}}_c},c = 1,2, \cdots ,{{MNJ} \mathord{\left/
 {\vphantom {{MNJ} K}} \right.
 \kern-\nulldelimiterspace} K}$.
We approximate the messages updated and passed between the factor node ${\bf{y}}$ and variable nodes ${{\bf{x}}_c},c = 1,2, \cdots ,{{MNJ} \mathord{\left/
 {\vphantom {{MNJ} K}} \right.
 \kern-\nulldelimiterspace} K}$ on the factor graph as Gaussian.
A detailed implementation of the Memory AMP detector is summarized in \textbf{Algorithm \ref{alg:A}}, and the steps of the $t$-th iteration are described as below:
\begin{algorithm}%[h]
\caption{Memory AMP Detector}
\label{alg:A}
\begin{algorithmic}
\STATE {Input: ${\bf{y}}$, ${\bf{H}}$, ${{\lambda _{\min }}}$, ${{\lambda _{\max }}}$, $L$ and $\mathcal{T}$.}
\STATE {Initialization: ${{\bm{\mu }}^{(1)}} = {{{\bf{\bar r}}}^{(0)}} = {\bf{0}}$, ${\lambda ^ + } = {{\left( {{\lambda _{\max }} + {\lambda _{\min }}} \right)} \mathord{\left/
 {\vphantom {{\left( {{\lambda _{\max }} + {\lambda _{\min }}} \right)} 2}} \right.
 \kern-\nulldelimiterspace} 2}$, ${\eta _{1,1}} = {{\left( {\frac{1}{{\cal N}}{{\bf{y}}^H}{\bf{y}} - \frac{{{\cal M}{\sigma ^2}}}{{\cal N}}} \right)} \mathord{\left/
 {\vphantom {{\left( {\frac{1}{{\cal N}}{{\bf{y}}^H}{\bf{y}} - \frac{{{\cal M}{\sigma ^2}}}{{\cal N}}} \right)} {{a_0}}}} \right.
 \kern-\nulldelimiterspace} {{a_0}}}$ and iteration count $t=1$.}
\REPEAT
\STATE {1)\; Factor node ${\bf{y}}$ generates the extrinsic mean ${\bf{r}}_c^{(t)}$ from (\ref{fac_mean}) and the extrinsic variance ${{\tau _{t,t}}\left( {\xi _t^*} \right)}$ from (\ref{fac_variance}), then delivers them to the variable nodes ${{\bf{x}}_c},c = 1,2, \cdots ,{{MNJ} \mathord{\left/
 {\vphantom {{MNJ} K}} \right.
 \kern-\nulldelimiterspace} K}$;}
\STATE {2)\; The variable nodes ${{\bf{x}}_c},c = 1,2, \cdots ,{{MNJ} \mathord{\left/
 {\vphantom {{MNJ} K}} \right.
 \kern-\nulldelimiterspace} K}$ calculate the mean vector ${{\bm{\mu }}^{(t + 1)}}$ in (\ref{var_mean}) and the variances ${\eta _{t + 1,t'}},1 \le t' \le t+1$ in (\ref{var_variance}), and passes them back to the factor node ${\bf{y}}$;}
\STATE {3)\; $t: = t + 1$;}
\UNTIL{desired convergence or $t=\mathcal{T}$.}
\STATE {Output: The decisions of the information bits for each user.}
\end{algorithmic}
\end{algorithm}

1) From factor node ${\bf{y}}$ to variable nodes ${{\bf{x}}_c},c = 1,2, \cdots ,{{MNJ} \mathord{\left/
 {\vphantom {{MNJ} K}} \right.
 \kern-\nulldelimiterspace} K}$: At the factor node ${\bf{y}}$, we can apply the LMMSE criterion to obtain the {\em a posteriori} estimate of ${\bf{x}}$, given by
\begin{align}\label{relation_LMMSE}
{{{\bf{\bar z}}}^{(t)}} = {{\bm{\mu }}^{(t)}} + {{\bf{H}}^H}{\left( {{\rho _t}{\bf{I}} + {\bf{H}}{{\bf{H}}^H}} \right)^{ - 1}}\left( {{\bf{y}} - {\bf{H}}{{\bm{\mu }}^{(t)}}} \right),
\end{align}
where ${\rho _t} = {{{\sigma ^2}} \mathord{\left/
 {\vphantom {{{\sigma ^2}} {{\eta _{t,t}}}}} \right.
 \kern-\nulldelimiterspace} {{\eta _{t,t}}}}$.
${{\bm{\mu }}^{(t)}} \in {\mathbb{C}^{\frac{{MNJD}}{K} \times 1}}$ and ${{\eta _{t,t}}}$ are the mean vector and variance received from variable nodes in the $(t-1)$-th iteration.

To avoid the large complexity of matrix inverse in (\ref{relation_LMMSE}), we introduce the following Lemmas for simplicity.
\begin{lemma}\label{Lem_1}
Assume that the matrix $({\bf{I}} - {\bf{C}})$ is invertible and the spectral radius of ${\bf{C}}$ satisfies $\rho ({\bf{C}}) < 1$. Then, ${\left( {{\bf{I}} - {\bf{C}}} \right)^{ - 1}} = \mathop {\lim }\limits_{t \to \infty } \sum\limits_{i = 0}^t {{{\bf{C}}^i}}$.
\end{lemma}
\begin{lemma}\label{Lem_2}
Starting with $t=1$ and ${{\bf{r}}^{(0)}} = {\bf{0}}$, we can use the recursive process ${{\bf{r}}^{(t)}} = {\bf{C}}{{\bf{r}}^{(t - 1)}} + {\bf{x}}$ to approximate $\sum\limits_{i = 0}^t {{{\bf{C}}^i}} {\bf{x}}\mathop  \to \limits^{t \to \infty } {\left( {{\bf{I}} - {\bf{C}}} \right)^{ - 1}}{\bf{x}}$.
\end{lemma}

Inspired by {\bf Lemma} \ref{Lem_1} and {\bf Lemma} \ref{Lem_2}, we define 
\begin{align}\label{relation_inv}
{{{\bf{\bar r}}}^{(t)}} = \left[ {{\bf{I}} - {\theta _t}\left( {{\rho _t}{\bf{I}} + {\bf{H}}{{\bf{H}}^H}} \right)} \right]{{{\bf{\bar r}}}^{(t - 1)}} + {\xi _t}( {{\bf{y}} - {\bf{H}}{{\bm{\mu }}^{(t)}}}),
\end{align}
where ${{\theta _t}}$ is a relaxation parameter to guarantee that the spectral radius of $\left[ {{\bf{I}} - {\theta _t}\left( {{\rho _t}{\bf{I}} + {\bf{H}}{{\bf{H}}^H}} \right)} \right]$ is less than 1. It is verified that ${\theta _t} = {\left( {{\lambda ^ + } + {\rho _t}} \right)^{ - 1}}$ with ${\lambda ^ + } = {{\left( {{\lambda _{\max }} + {\lambda _{\min }}} \right)} \mathord{\left/
 {\vphantom {{\left( {{\lambda _{\max }} + {\lambda _{\min }}} \right)} 2}} \right.
 \kern-\nulldelimiterspace} 2}$ satisfies such condition, where ${{\lambda _{\max }}}$ and ${{\lambda _{\min }}}$ are the maximal and minimal eigenvalues of ${{\bf{H}}{{\bf{H}}^H}}$, respectively. In addition, the weight ${\xi _t}$ is chosen and optimized to accelerate the convergence of Memory AMP detector. For convenience, we define ${\bf{B}} = {\lambda ^ + }{\bf{I}} - {\bf{H}}{{\bf{H}}^H}$ and yield ${\theta _t}{\bf{B}} = {\bf{I}} - {\theta _t}\left( {{\rho _t}{\bf{I}} + {\bf{H}}{{\bf{H}}^H}} \right)$.
\begin{remark}
Note that the complexity of calculating ${{\lambda _{\max }}}$ and ${{\lambda _{\min }}}$ is as high as that of the matrix inverse. Fortunately, a simple bound approximations of maximum and minimum eigenvalues can be applied without performance loss \cite{9805776}. 
\end{remark}

Therefore, starting with $t=1$ and ${{\bm{\mu }}^{(1)}} = {{{\bf{\bar r}}}^{(0)}} = {\bf{0}}$, we can approximately rewrite (\ref{relation_LMMSE}) as
\begin{subequations}
\begin{align}
&{{\bf{z}}^{(t)}} = {{\bm{\mu }}^{(t)}} + \,{{\bf{H}}^H}{{{\bf{\bar r}}}^{(t)}}\label{fac_rec}\\
 &= \underbrace {\sum\limits_{i = 1}^t {{\phi _{t,i}}{{\bf{H}}^H}{{\bf{B}}^{t - i}}} }_{{{\bf{Q}}_t}}{\bf{y}} + \sum\limits_{i = 1}^t {\underbrace {\left( { - {\phi _{t,i}}{{\bf{A}}_{t - i}}} \right)}_{{{\bf{F}}_{t,i}}}{{\bm{\mu }}^{(i)}}}  + {{\bm{\mu }}^{(t)}},\label{recurse_ite}
\end{align}
\end{subequations}
where (\ref{recurse_ite}) follows the recursive process of (\ref{relation_inv}), and we define ${{\bf{A}}_t} = {{\bf{H}}^H}{{\bf{B}}^t}{\bf{H}}$ and $\begin{aligned}
{{\phi _{t,i}}} 
=
\begin{cases}
{\xi _t},&i=t\\
{\xi _i}\prod\nolimits_{j = i + 1}^t {{\theta _j}}, &i<t
\end{cases}.
\end{aligned}$
%\begin{equation*}
%\begin{aligned}
%{{\phi _{t,i}}} 
%=
%\begin{cases}
%{\xi _t},&i=t\\
%{\xi _i}\prod\nolimits_{j = i + 1}^t {{\theta _j}}, &i<t
%\end{cases}.
%\end{aligned}
%\end{equation*}

From (\ref{recurse_ite}), we notice that all the preceding messages $\left\{ {{{\bm{\mu }}^{(t)}}} \right\}$ are utilized for estimation. Thus, the traditional orthogonality principle between the current input and output estimation errors in non-memory OAMP/VAMP \cite{ge2021otfs} and EP \cite{9827946} is not sufficient to guarantee the asymptotic independent identically distributed (IID) Gaussianity of estimation errors in Memory AMP. Alternatively, a stricter orthogonality is required, i.e., the current output estimation error is required to be orthogonal to all preceding input estimation errors.
Following the orthogonalization rule \cite{9805776}, we then generate the extrinsic mean vector of the estimation, given by
\begin{align}\label{fac_mean}
{{\bf{r}}^{(t)}} = \frac{1}{{{\varepsilon _t}}}\left( {{{\bf{z}}^{(t)}} - \sum\limits_{i = 1}^t {{{c}_{t,i}}{{\bm{\mu }}^{(i)}}} } \right),
\end{align}
where ${\varepsilon _t} = 1 - \sum\limits_{i = 1}^t {{{c}_{t,i}}}$, ${a_t} = \frac{1}{N}\text{tr}\{ {{\bf{A}}_t}\}$ and $\begin{aligned}
{{{c}_{t,i}}} 
=
\begin{cases}
1 - {\phi _{t,t}}{a_0},&i=t\\
- {\phi _{t,i}}{a_{t - i}}, &i<t
\end{cases}.
\end{aligned}$
%\begin{proof}
%Following the orthogonalization rule \cite{9805776}, we have
%\begin{align*}
%{{c}_{t,i}} = \frac{1}{N}\text{tr}\{ {{\bf{F}}_{t,i}}\}  =  - \frac{1}{N}\text{tr}\{ {\phi _{t,i}}{{\bf{A}}_{t - i}}\}  =  - {\phi _{t,i}}{a_{t - i}}
%\end{align*}
%for $i<t$. In a similar fashion, we have
%\begin{align*}
%{{c}_{t,t}} = \frac{1}{N}\text{tr}\{ {\bf{I}} + {{\bf{F}}_{t,t}}\}  = \frac{1}{N}\text{tr}\{ {\bf{I}} - {\phi _{t,t}}{{\bf{A}}_0}\}  = 1 - {\phi _{t,t}}{a_0}
%\end{align*}
%for $i=t$. We also calculate
%\begin{align*}
%{\varepsilon _t} = \frac{1}{N}\text{tr}\{ {{\bf{Q}}_t}{\bf{H}}\}  = \frac{1}{N}\text{tr}\{ \sum\limits_{i = 1}^t {{\phi _{t,i}}{{\bf{H}}^H}{{\bf{B}}^{t - i}}} {\bf{H}}\}  = \sum\limits_{i = 1}^t {{\phi _{t,i}}\frac{1}{N}\text{tr}\{ {{\bf{A}}_{t - i}}\} }  = \sum\limits_{i = 1}^t {{\phi _{t,i}}{a_{t - i}}}  = 1 - \sum\limits_{i = 1}^t {{{c}_{t,i}}},
%\end{align*}
%which completes the proof.
%\end{proof}

We can further denote the extrinsic mean vector ${{\bf{r}}^{(t)}}$ as ${{\bf{r}}^{(t)}} = {\left[ {{{\left( {{\bf{r}}_1^{(t)}} \right)}^T},{{\left( {{\bf{r}}_2^{(t)}} \right)}^T}, \cdots ,{{\left( {{\bf{r}}_{{{MNJ} \mathord{\left/
 {\vphantom {{MNJ} K}} \right.
 \kern-\nulldelimiterspace} K}}^{(t)}} \right)}^T}} \right]^T}\in {\mathbb{C}^{\frac{{MNJD}}{K} \times 1}}$, where ${\bf{r}}_c^{(t)}\in {\mathbb{C}^{D \times 1}}$.
Next, the extrinsic variance of the estimation can be approximately calculated by
\begin{subequations}\label{factor_var}
\begin{align}
{\tau _{t,t}} &= \frac{1}{\mathcal{N}}\mathbb{E}\left\{ {\left\| {{{\bf{r}}^{(t)}} - {\bf{x}}} \right\|_2^2} \right\}\\
%& = \frac{1}{{\cal N}}\mathbb{E}\left\{ {\left\| {\frac{1}{{{\varepsilon _t}}}\left[ {{{\bf{Q}}_t}{\bf{y}} + \sum\limits_{i = 1}^t {{{\bf{E}}_{t,i}}{{\bm{\mu }}^{(i)}}} } \right] - {\bf{x}}} \right\|_2^2} \right\}\\
& = \frac{1}{{\cal N}}\mathbb{E}\left\{ {\left\| {\frac{1}{{{\varepsilon _t}}}\left( {{{\bf{Q}}_t}{\bm{\omega }} + \sum\limits_{i = 1}^t {{{\bf{E}}_{t,i}}{{\bf{f}}^{(i)}}} } \right)} \right\|_2^2} \right\},
\end{align}
\end{subequations}
where $\mathbb{E}\{  \cdot \} $ denotes the expectation operation and ${{\bf{E}}_{t,i}} = {\phi _{t,i}}\left( {{a_{t - i}}{\bf{I}} - {{\bf{A}}_{t - i}}} \right)$. We also define the estimation error of ${{\bm{\mu }}^{(i)}}$ as ${{\bf{f}}^{(i)}} = {{\bm{\mu }}^{(i)}} - {\bf{x}}$, where $\mathbb{E}\left\{ {{{\bf{f}}^{(j)}}{{\left( {{{\bf{f}}^{(i)}}} \right)}^H}} \right\} = {\eta _{i,j}}{\bf{I}}$.

Assuming the noise vector $\bm{\omega }$ is independent with $\left\{ {{{\bf{f}}^{(i)}}} \right\}$, we can further simplify (\ref{factor_var}) as
\begin{align}
{\tau _{t,t}} 
%&= \mathop {\lim }\limits_{{\cal N} \to \infty } \frac{1}{{{\cal N}\varepsilon _t^2}}{\left( {{{\bf{Q}}_t}{\bm{\omega }} + \sum\limits_{i = 1}^t {{{\bf{E}}_{t,i}}{{\bf{f}}^{(i)}}} } \right)^H}\left( {{{\bf{Q}}_t}{\bm{\omega }} + \sum\limits_{i = 1}^t {{{\bf{E}}_{t,i}}{{\bf{f}}^{(i)}}} } \right)\nonumber\\
& = \frac{1}{{\varepsilon _t^2}}\left[ {{\sigma ^2}\frac{1}{{\cal N}}\text{tr}\left\{ {{\bf{Q}}_t^H{{\bf{Q}}_t}} \right\} + \sum\limits_{i = 1}^t {\sum\limits_{j = 1}^t {{\eta _{i,j}}\frac{1}{{\cal N}}\text{tr}\left\{ {{\bf{E}}_{t,i}^H{{\bf{E}}_{t,j}}} \right\}} } } \right]\nonumber\\
& = \frac{1}{{\varepsilon _t^2}}\sum\limits_{i = 1}^t {\sum\limits_{j = 1}^t {{\phi _{t,i}}{\phi _{t,j}}\left( {{\sigma ^2}{a_{2t - i - j}} + {\eta _{i,j}}{{\bar a}_{t - i,t - j}}} \right)} }\nonumber \\
& = \frac{{{\varpi _{t,1}}\xi _t^2 - 2{\varpi _{t,2}}{\xi _t} + {\varpi _{t,3}}}}{{a_0^2{{\left( {{\varpi _{t,0}} + {\xi _t}} \right)}^2}}}.\label{fac_variance}
\end{align}
Here, we have used the following notations:
${{\bar a}_{i,j}} = {\lambda ^ + }{a_{i + j}} - {a_{i + j + 1}} - {a_i}{a_j}$, ${\varpi _{t,0}} =  - {{\sum\nolimits_{i = 1}^{t - 1} {{{c}_{t,i}}} } \mathord{\left/
 {\vphantom {{\sum\nolimits_{i = 1}^{t - 1} {{{c}_{t,i}}} } {{a_0}}}} \right.
 \kern-\nulldelimiterspace} {{a_0}}}$, ${\varpi _{t,1}} = {\sigma ^2}{a_0} + {\eta _{t,t}}{{\bar a}_{0,0}}$,
${\varpi _{t,2}} =  - \sum\nolimits_{i = 1}^{t - 1} {{\phi _{t,i}}\left( {{\sigma ^2}{a_{t - i}} + {\eta _{t,i}}{{\bar a}_{0,t - i}}} \right)}$ and ${\varpi _{t,3}} = \sum\nolimits_{i = 1}^{t - 1} {\sum\nolimits_{j = 1}^{t - 1} {{\phi _{t,i}}{\phi _{t,j}}\left( {{\sigma ^2}{a_{2t - i - j}} + {\eta _{i,j}}{{\bar a}_{t - i,t - j}}} \right)} } $.

The optimal parameter ${{\xi _t}}$ is obtained by minimizing ${\tau _{t,t}}$. Since ${\tau _{t,t}}\left( {{\xi _t}} \right)$ is differentiable with respect to ${{\xi _t}}$ except at the point ${\xi _t} =  - {\varpi _{t,0}}$, but ${\tau _{t,t}}\left( { - {\varpi _{t,0}}} \right) =  + \infty$. Therefore, the optimal ${{\xi _t}}$ is either $\pm \infty $ or ${{\partial {\tau _{t,t}}\left( {{\xi _t}} \right)} \mathord{\left/
 {\vphantom {{\partial {\tau _{t,t}}\left( {{\xi _t}} \right)} {\partial {\xi _t}}}} \right.
 \kern-\nulldelimiterspace} {\partial {\xi _t}}} = 0$. As a result, we set the optimal solution $\xi _1^* = 1$ and for $t \ge 2$,
\begin{equation}
\begin{aligned}
\xi _t^*
=
\begin{cases}
\frac{{{\varpi _{t,0}}{\varpi _{t,2}} + {\varpi _{t,3}}}}{{{\varpi _{t,0}}{\varpi _{t,1}} + {\varpi _{t,2}}}},&{\varpi _{t,0}}{\varpi _{t,1}} + {\varpi _{t,2}} \ne 0\\
+\infty , &\text{otherwise}
\end{cases}.
\end{aligned}
\end{equation} 

The extrinsic mean ${\bf{r}}_c^{(t)}$ and variance ${{\tau _{t,t}}\left( {\xi _t^*} \right)}$ are finally delivered to variable nodes ${{\bf{x}}_c},c = 1,2, \cdots ,{{MNJ} \mathord{\left/
 {\vphantom {{MNJ} K}} \right.
 \kern-\nulldelimiterspace} K}$.

2) From variable nodes ${{\bf{x}}_c},c = 1,2, \cdots ,{{MNJ} \mathord{\left/
 {\vphantom {{MNJ} K}} \right.
 \kern-\nulldelimiterspace} K}$ to factor node ${\bf{y}}$: 
At each variable node, we can express the {\em a posteriori} probability as
\begin{align}\label{var_post}
{{\bar P}^{(t)}}( {{{\bf{x}}_c} \!=\! {{\bm{\chi }}_j}} ) \!\propto\! {P_D}( {{{\bf{x}}_c}\! =\! {{\bm{\chi }}_j}} )\exp \left( { - \frac{{\| {{{\bm{\chi }}_j} - {\bf{r}}_c^{(t)}} \|_2^2}}{{{\tau _{t,t}}\left( {\xi _t^*} \right)}}} \right),
\end{align}
where $\forall {{\bm{\chi }}_j} \in {{\bar {\mathbb{A}}}_j}$, $j = \left\lceil {\frac{{cK}}{{MN}}} \right\rceil$ and $\left\lceil  \cdot \right\rceil $ denotes the round up operation. ${{\bar {\mathbb{A}}}_j}$ is a set contains the non-zero elements of $\mathbb{A}_j$, and ${{\bm{\chi }}_j}$ is a $D$-dimensional codeword from ${{\bar {\mathbb{A}}}_j}$. ${P_D}\left( {{{\bf{x}}_c} = {{\bm{\chi }}_j}} \right)$ denotes the {\em a priori} probability and is assumed to be equiprobable symbols if no prior information observed. The {\em a posteriori} probability is then projected into Gaussian distributions ${\cal C}{\cal N}\left( {g_c^{(t)}[i],\delta _c^{(t)}[i]} \right), i = 1,2, \cdots ,D$, with 
\begin{subequations}\label{var_Gaus_pro}
\begin{align}
g_c^{(t)}[i] = \sum\limits_{{{\bm{\chi }}_j} \in {{\bar {\mathbb{A}}}_j}} {{{\bar P}^{(t)}}\left( {{{\bf{x}}_c} = {{\bm{\chi }}_j}} \right){\chi _j}[i]},
\end{align}
\begin{align}
\delta _c^{(t)}[i] = \sum\limits_{{{\bm{\chi }}_j} \in {{\bar {\mathbb{A}}}_j}} {{{\bar P}^{(t)}}\left( {{{\bf{x}}_c} = {{\bm{\chi }}_j}} \right){{\left| {{\chi _j}[i]} \right|}^2} - {{\left| {g_c^{(t)}[i]} \right|}^2}}.
\end{align}
\end{subequations}
For simplicity, we further do the sample average of the variance, i.e., $\delta  = \frac{1}{{\cal N}}\sum\limits_{c = 1}^{{{MNJ} \mathord{\left/
 {\vphantom {{MNJ} K}} \right.
 \kern-\nulldelimiterspace} K}} {\sum\limits_{i = 1}^D {\delta _c^{(t)}[i]} }$. Following the Gaussian message combing rule \cite{ge2021otfs}, we then update the extrinsic variance and mean, given by
\begin{subequations}\label{var_Gau_cancel}
\begin{align}
{{\bar \eta }_{t + 1,t + 1}} = {\left[ {{{\left( \delta  \right)}^{ - 1}} - {{\left( {{\tau _{t,t}}\left( {\xi _t^*} \right)} \right)}^{ - 1}}} \right]^{ - 1}},
\end{align}
\begin{align}
\bar \mu _c^{(t + 1)}[i] = {{\bar \eta }_{t + 1,t + 1}}\left[ {\frac{{g_c^{(t)}[i]}}{\delta } - \frac{{r_c^{(t)}[i]}}{{{\tau _{t,t}}\left( {\xi _t^*} \right)}}} \right].
\end{align}
\end{subequations}
As a result, 
${{{\bm{\bar \mu }}}^{(t + 1)}} = \left[ {{{\left( {{\bm{\bar \mu }}_1^{(t + 1)}} \right)}^T},{{\left( {{\bm{\bar \mu }}_2^{(t + 1)}} \right)}^T}, \cdots ,} \right.$ ${\left. {{{\left( {{\bm{\bar \mu }}_{{{MNJ} \mathord{\left/
 {\vphantom {{MNJ} K}} \right.
 \kern-\nulldelimiterspace} K}}^{(t + 1)}} \right)}^T}} \right]^T}\in {\mathbb{C}^{\frac{{MNJD}}{K} \times 1}}$.
%${{{\bm{\bar \mu }}}^{(t + 1)}} = {\left[ {{{\left( {{\bm{\bar \mu }}_1^{(t + 1)}} \right)}^T},{{\left( {{\bm{\bar \mu }}_2^{(t + 1)}} \right)}^T}, \cdots ,{{\left( {{\bm{\bar \mu }}_{{{MNJ} \mathord{\left/
% {\vphantom {{MNJ} K}} \right.
% \kern-\nulldelimiterspace} K}}^{(t + 1)}} \right)}^T}} \right]^T}\in {\mathbb{C}^{\frac{{MNJD}}{K} \times 1}}$.

To guarantee the convergence and improve the performance of the detector algorithm, we apply a damping vector ${{\bf{\Lambda }}_{t + 1}} = {\left[ {{\Lambda _{t + 1,1}},{\Lambda _{t + 1,2}}, \cdots ,{\Lambda _{t + 1,t + 1}}} \right]^T}$ under the constraint of $\sum\limits_{i = 1}^{t + 1} {{\Lambda _{t + 1,i}}}  = 1$. Therefore, the variable nodes further update the mean vector ${{\bm{\mu }}^{(t + 1)}}$ as
\begin{align}\label{var_mean}
{{\bm{\mu }}^{(t + 1)}} = \left[ {{{\bm{\mu }}^{(1)}},{{\bm{\mu }}^{(2)}}, \cdots ,{{\bm{\mu }}^{(t)}},{{{\bm{\bar \mu }}}^{(t + 1)}}} \right] \cdot {{\bf{\Lambda }}_{t + 1}}.
\end{align}
Next, the extrinsic variance can be approximately updated by
\begin{align}\label{variable_var}
{\eta _{t + 1,t + 1}} = \frac{1}{{\cal N}}\mathbb{E}\left\{ {\| {{{\bm{\mu }}^{(t + 1)}} - {\bf{x}}} \|_2^2} \right\} = {\bf{\Lambda }}_{t + 1}^H{{{\bf{\bar V}}}_{t + 1}}{{\bf{\Lambda }}_{t + 1}},
\end{align}
where ${{{\bf{\bar V}}}_{t + 1}} = {\left[ {\begin{array}{*{20}{c}}
{{{\bf{V}}_t}}&{\begin{array}{*{20}{c}}
{{{\bar \eta }_{1,t + 1}}}\\
 \vdots 
\end{array}}\\
{\begin{array}{*{20}{c}}
{{{\bar \eta }_{t + 1,1}}}& \cdots 
\end{array}}&{{{\bar \eta }_{t + 1,t + 1}}}
\end{array}} \right]_{(t + 1) \times (t + 1)}}$ with ${{\bf{V}}_t} = {\left[ {{\eta _{i,j}}} \right]_{t \times t}},1 \le i \le j \le t$. For $1 \le t' \le t$, we can calculate \cite{9805776}
\begin{align}
&{{\bar \eta }_{t + 1,t'}} = \frac{1}{{\cal N}}\mathbb{E}\left\{ {{{\left[ {{{{\bm{\bar \mu }}}^{(t + 1)}} - {\bf{x}}} \right]}^H}{{\bf{f}}^{(t')}}} \right\}\nonumber\\
&\!\approx\! \mathop {\lim }\limits_{{\cal N} \to \infty } {{\left[ {\frac{1}{{\cal N}}{{\left( {{\bf{y}} \!-\! {\bf{H}}{{{\bm{\bar \mu }}}^{(t + 1)}}} \right)}^H}\left( {{\bf{y}} \!-\! {\bf{H}}{{\bm{\mu }}^{(t')}}} \right) \!-\! \frac{{\mathcal{M}{\sigma ^2}}}{{\cal N}}} \right]} \mathord{\left/
 {\vphantom {{\left[ {\frac{1}{{\cal N}}{{\left( {{\bf{y}} - {\bf{H}}{{{\bm{\bar \mu }}}^{(t + 1)}}} \right)}^H}\left( {{\bf{y}} - {\bf{H}}{{\bm{\mu }}^{(t')}}} \right) - \frac{{\mathcal{M}{\sigma ^2}}}{{\cal N}}} \right]} {{a_0}}}} \right.
 \kern-\nulldelimiterspace} {{a_0}}},\label{var_dam_var}
\end{align}
where ${{\bar \eta }_{t',t + 1}}$ is equal to the conjugate of ${{\bar \eta }_{t + 1,t'}}$.
%\begin{proof}
%See Appendix.
%\end{proof}

Different from the heuristic damping method in the literature \cite{raviteja2018interference,xiang2021gaussian,9827946}, here, we solve the following optimization problem based on (\ref{variable_var}) to obtain the solution of damping vector ${{\bf{\Lambda }}_{t + 1}}$,
\begin{subequations}\label{variab_opt}
\begin{align}\mathop {\min }\limits_{{{\bf{\Lambda }}_{t + 1}}}\quad &\mathop \frac{1}{2}{\bf{\Lambda }}_{t + 1}^H{{{\bf{\bar V}}}_{t + 1}}{{\bf{\Lambda }}_{t + 1}}\\
\text{s.t.}\quad &{{\bf{1}}^T}{{\bf{\Lambda }}_{t + 1}} = 1
\end{align}
\end{subequations}
where ${\bf{1}}$ is an all one vector. As ${{{\bf{\bar V}}}_{t + 1}}$ is a positive semi-definite matrix in general, problem (\ref{variab_opt}) is a convex optimization problem and can be easily solved. It is verified that the optimal solution is given by
\begin{equation}\label{damping_solu}
\begin{aligned}
{\bf{\Lambda }}_{t + 1}^*
=
\begin{cases}
\frac{{{{\left( {{{{\bf{\bar V}}}_{t + 1}}} \right)}^{ - 1}}{\bf{1}}}}{{{{\bf{1}}^T}{{\left( {{{{\bf{\bar V}}}_{t + 1}}} \right)}^{ - 1}}{\bf{1}}}},&\text{if}\; {{{{\bf{\bar V}}}_{t + 1}}}\; \text{is invertible}\\
{\left[ {0,0, \cdots ,1,0} \right]^T}, &\text{otherwise}
\end{cases}.
\end{aligned}
\end{equation} 
Following (\ref{damping_solu}), we can update the variance straightforwardly as following
\begin{equation}\label{var_variance}
\begin{aligned}
{\eta _{t',t + 1}} = &{\eta _{t + 1,t'}} = {\eta _{t + 1,t + 1}}
\\&=
\begin{cases}
\frac{1}{{{{\bf{1}}^T}{{\left( {{{{\bf{\bar V}}}_{t + 1}}} \right)}^{ - 1}}{\bf{1}}}},&\text{if}\; {{{{\bf{\bar V}}}_{t + 1}}}\; \text{is invertible}\\
{\eta _{t,t}}, &\text{otherwise}
\end{cases}
\end{aligned}
\end{equation} 
for $1 \le t' \le t$.

Finally, ${{\bm{\mu }}^{(t + 1)}}$ and ${\eta _{t + 1,t'}},1 \le t' \le t+1$ are passed back to the factor node.
\begin{remark}
In general, we consider a maximum damping length $L$ (i.e., the number of non-zero entries in ${{\bf{\Lambda }}_{t + 1}}$) instead of full damping, where $L=3$ or $2$ is sufficient for desired performance.
\end{remark}

3) Stopping Criterion: The Memory AMP detector stops until the desired convergence or the maximum iteration number $\mathcal{T}$ is reached. 

Finally, we make decisions of the transmitted symbols and apply the SCMA demapping to recover the transmitted information bits of each user. 
%The structure of the proposed Memory AMP detector is shown in Fig. \ref{fig2_structure}.
%\begin{figure}
%  \centering
%  \includegraphics[width=4in]{fig2_structure.pdf}
%  \caption{Structure of the proposed Memory AMP detector.}
%  \label{fig2_structure}
%\end{figure}

\subsection{Complexity Analysis}

As we can see, the equivalent channel matrix ${\bf{H}}$ is a sparse matrix and only matrix-vector multiplications are involved in the Memory AMP detector. Thus, our proposed Memory AMP detector has a relatively low complexity. Specifically, the complexity of Memory AMP in each iteration is mainly dominated by (\ref{relation_inv}), (\ref{fac_rec}), (\ref{var_post}), (\ref{var_Gaus_pro}), (\ref{var_Gau_cancel}) and (\ref{var_dam_var}), which require a complexity order $\mathcal{O}\left( {UMN({S_{\bf{B}}} + {S_{\bf{H}}})} \right)$, $\mathcal{O}\left( {UMN{S_{\bf{H}}}} \right)$, $\mathcal{O}\left( {\frac{{MNJDQ}}{K}} \right)$, $\mathcal{O}\left( {\frac{{2MNJDQ}}{K}} \right)$, $\mathcal{O}\left( {\frac{{2MNJD}}{K}} \right)$ and $\mathcal{O}\left( {UMN({S_{\bf{H}}} + 1)} \right)$, respectively. Here, we use ${{S_{\bf{B}}}}$ and ${{S_{\bf{H}}}}$ represent the average number of non-zero entries in each row of ${\bf{B}}$ and ${\bf{H}}$, respectively. Therefore, the overall computational complexity of Memory AMP is $\mathcal{O}\left( {\left( {UMN({S_{\bf{B}}} + 3{S_{\bf{H}}} + 1) + \frac{{MNJD(2 + 3Q)}}{K}} \right)\mathcal{T}} \right)$. To summarize, a detailed complexity comparison of GMP \cite{raviteja2018interference,xiang2021gaussian}, EP \cite{ge2023otfs,9827946}, OAMP/VAMP \cite{ge2021otfs} and our Memory AMP is illustrated in TABLE~\ref{tab1}. It is obvious that our proposed Memory AMP can achieve comparable complexity with GMP and EP, and has relatively low complexity than OAMP/VAMP.
\begin{table}%[h!]
  \begin{center}
    \caption{Complexity comparison between different detectors.}
    \label{tab1}
    \begin{tabular}{c|c} % <-- Alignments: 1st column left, 2nd middle and 3rd right, with vertical lines in between
    \hline
      Detectors & Computational Complexity\\
      \hline
      \thead{Memory\\AMP} & $\mathcal{O}\left( {\left( {UMN({S_{\bf{B}}} + 3{S_{\bf{H}}} + 1) + \frac{{MNJD(2 + 3Q)}}{K}} \right)\mathcal{T}} \right)$\\
      \hline
      EP & $\mathcal{O}\left( {\left( {UMN(6{S_{\bf{H}}}D + {S_{\bf{H}}}DQ) + \frac{{2MNJDQ}}{K}} \right)\mathcal{T}} \right)$\\
      \hline
      GMP & $\mathcal{O}\left( {\left( {UMN(3{S_{\bf{H}}}DQ + 2{S_{\bf{H}}})} \right)\mathcal{T}} \right)$\\
      \hline
      \thead{OAMP\\VAMP} & \thead{${\cal O}\left( {\left( {{{\left( {\frac{{MNJD}}{K}} \right)}^3} + {{\left( {\frac{{MNJD}}{K}} \right)}^2}} \right.} \right.$\\$\left. {\left. { + UMN(6{S_{\bf{H}}}D + {S_{\bf{H}}}DQ + {S_{\bf{H}}}) + \frac{{2MNJDQ}}{K}} \right){\cal T}} \right)$}\\
      \hline
    \end{tabular}
  \end{center}
  \vspace{-1.5em}
\end{table}

%\begin{table*}%[h!]
%  \begin{center}
%    \caption{Complexity comparison between different detectors.}
%    \label{tab1}
%    \begin{tabular}{c|c} % <-- Alignments: 1st column left, 2nd middle and 3rd right, with vertical lines in between
%    \hline
%      Detectors & Computational Complexity\\
%      \hline
%      Memory AMP & $\mathcal{O}\left( {\left( {UMN({S_{\bf{B}}} + 3{S_{\bf{H}}} + 1) + \frac{{MNJD(2 + 3Q)}}{K}} \right)\mathcal{T}} \right)$\\
%      EP & $\mathcal{O}\left( {\left( {UMN(6{S_{\bf{H}}}D + {S_{\bf{H}}}DQ) + \frac{{2MNJDQ}}{K}} \right)\mathcal{T}} \right)$\\
%      GMP & $\mathcal{O}\left( {\left( {UMN(3{S_{\bf{H}}}DQ + 2{S_{\bf{H}}})} \right)\mathcal{T}} \right)$\\
%      OAMP/VAMP & $\mathcal{O}\left( {\left( {{{\left( {\frac{{MNJD}}{K}} \right)}^3} + {{\left( {\frac{{MNJD}}{K}} \right)}^2} + UMN(6{S_{\bf{H}}}D + {S_{\bf{H}}}DQ + {S_{\bf{H}}}) + \frac{{2MNJDQ}}{K}} \right)\mathcal{T}} \right)$\\
%      \hline
%    \end{tabular}
%  \end{center}
%\end{table*}

\section{Simulation Results}\label{V_simulation}

In this section, we test the performance of the Memory AMP detector for MIMO-OTFS SCMA systems. We consider the carrier frequency is centered at $4$ GHz and subcarrier spacing $\Delta f=15$ kHz. The RRC rolloff factor is set to $0.4$ for both the transmitter and receiver. Unless otherwise stated, the delay-Doppler plane consists of $M=32$ and $N=16$. We assume that $J=6$ users are sharing $K=4$ orthogonal resources at the same time with $U=4$ receive antennas at the BS. We apply the SCMA codebooks according to \cite{xiao2018capacity} with size $Q=4$ and $D=2$ non-zero entries in each codeword. A typical urban channel model \cite{channel2017} is adopted with exponential power delay profile. The velocity of each user is set to $300$ km/h, resulting in a maximum Doppler frequency shift $\nu _{\text{max}} =1111$ Hz. We further generate the channel Doppler shift by using the Jakes formulation  \cite{raviteja2018interference,ge2021receiver}, i.e., $\nu _{uj,i} =\nu _{\text{max}}\cos \left( {{\varrho _{uj,i}}} \right),\forall u,j,i$, where ${{\varrho _{uj,i}}}$ is uniformly distributed over $\left[ { - \pi ,\pi } \right]$.
\begin{figure*}%[ht]
\begin{subfigure}{.33\textwidth}
  \centering
  % include first image
  \includegraphics[width=1\linewidth]{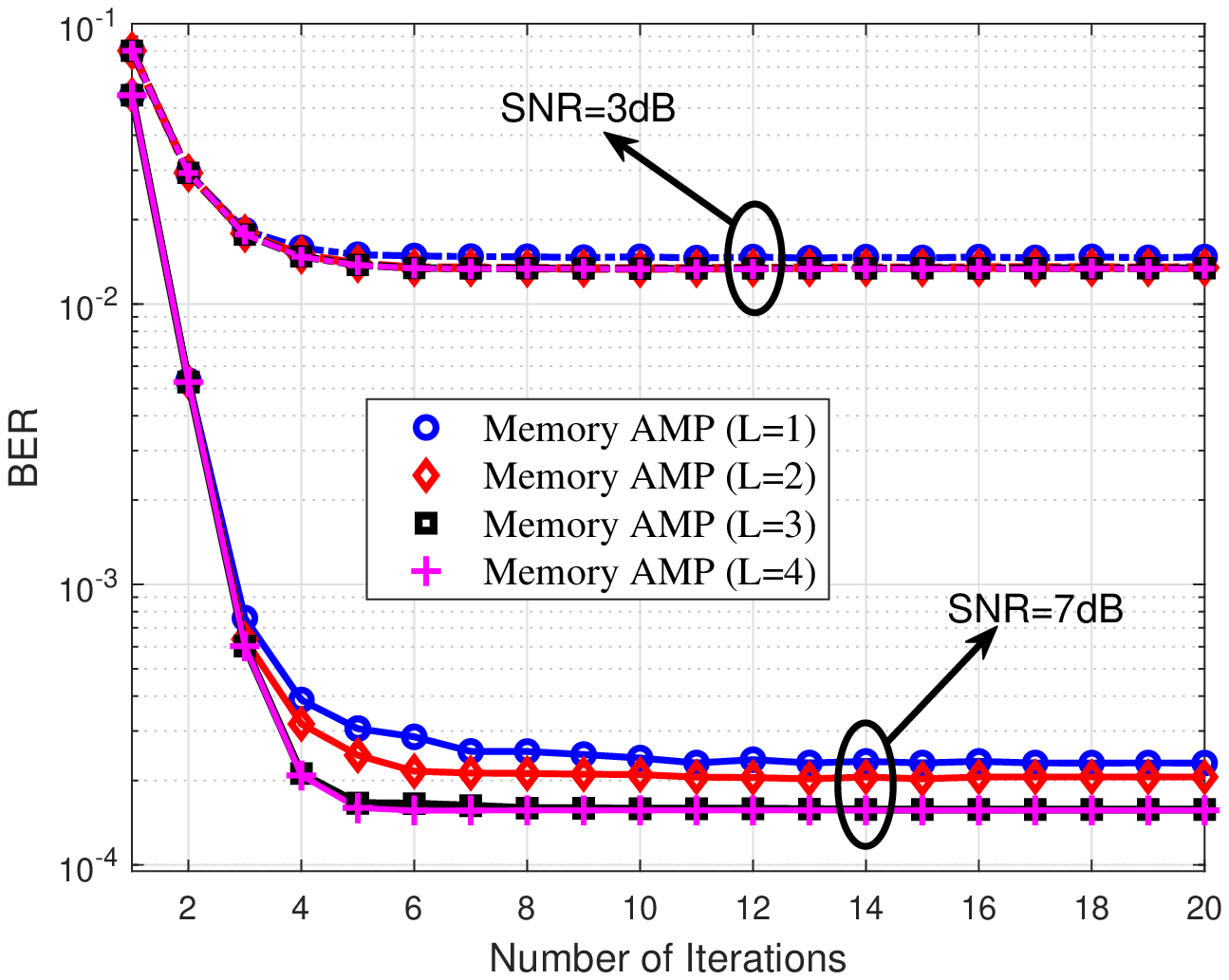}
  \caption{Convergence analysis of Memory AMP with different damping length.}
  \label{Convergence_damping}
\end{subfigure}
\begin{subfigure}{.33\textwidth}
  \centering
  % include second image
  \includegraphics[width=1\linewidth]{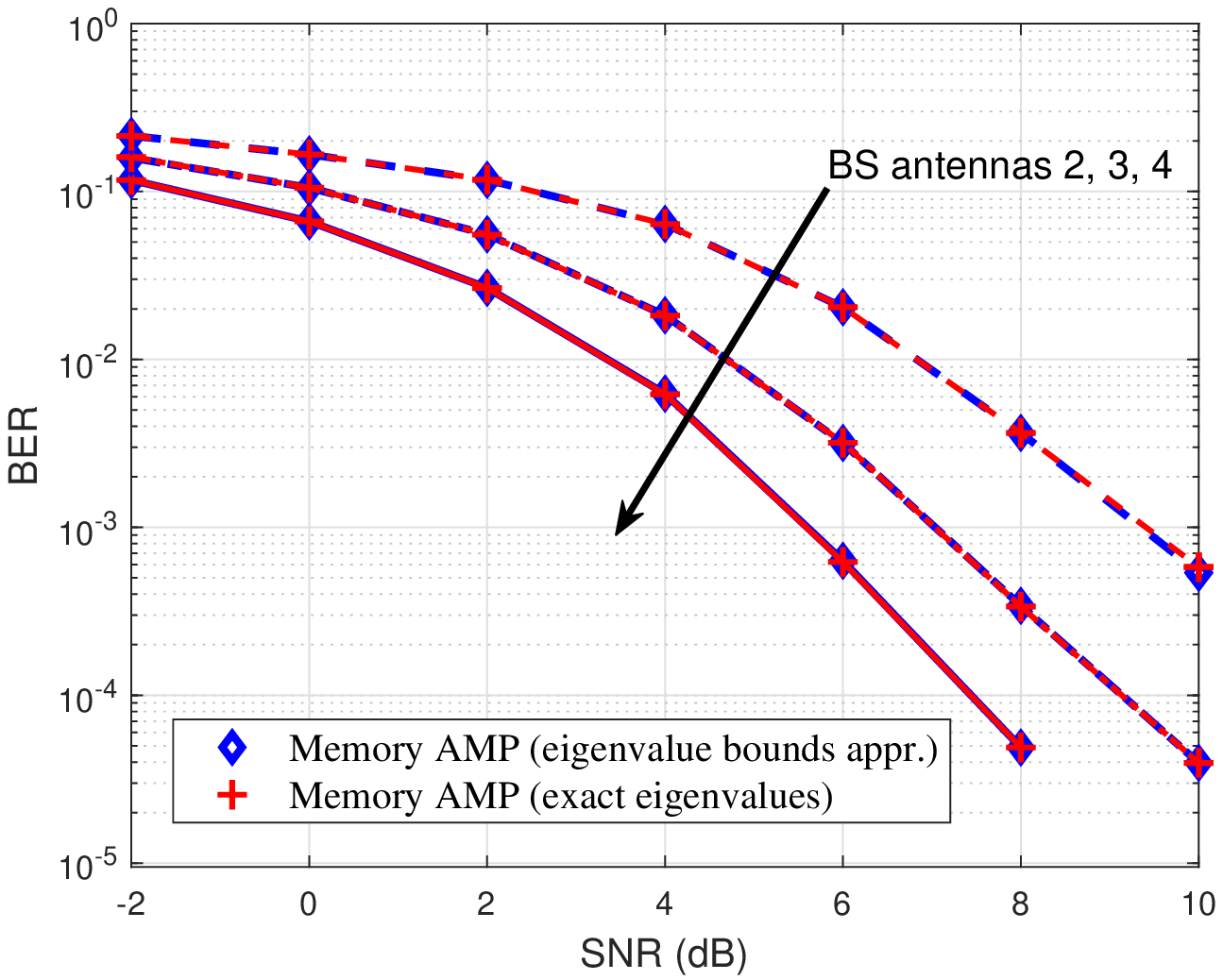}
  \caption{BER performance test of the eigenvalue bounds approximation for different number of BS antennas.}
  \label{antenna_eigenvalue}
\end{subfigure}
\begin{subfigure}{.33\textwidth}
  \centering
  % include third image
  \includegraphics[width=1\linewidth]{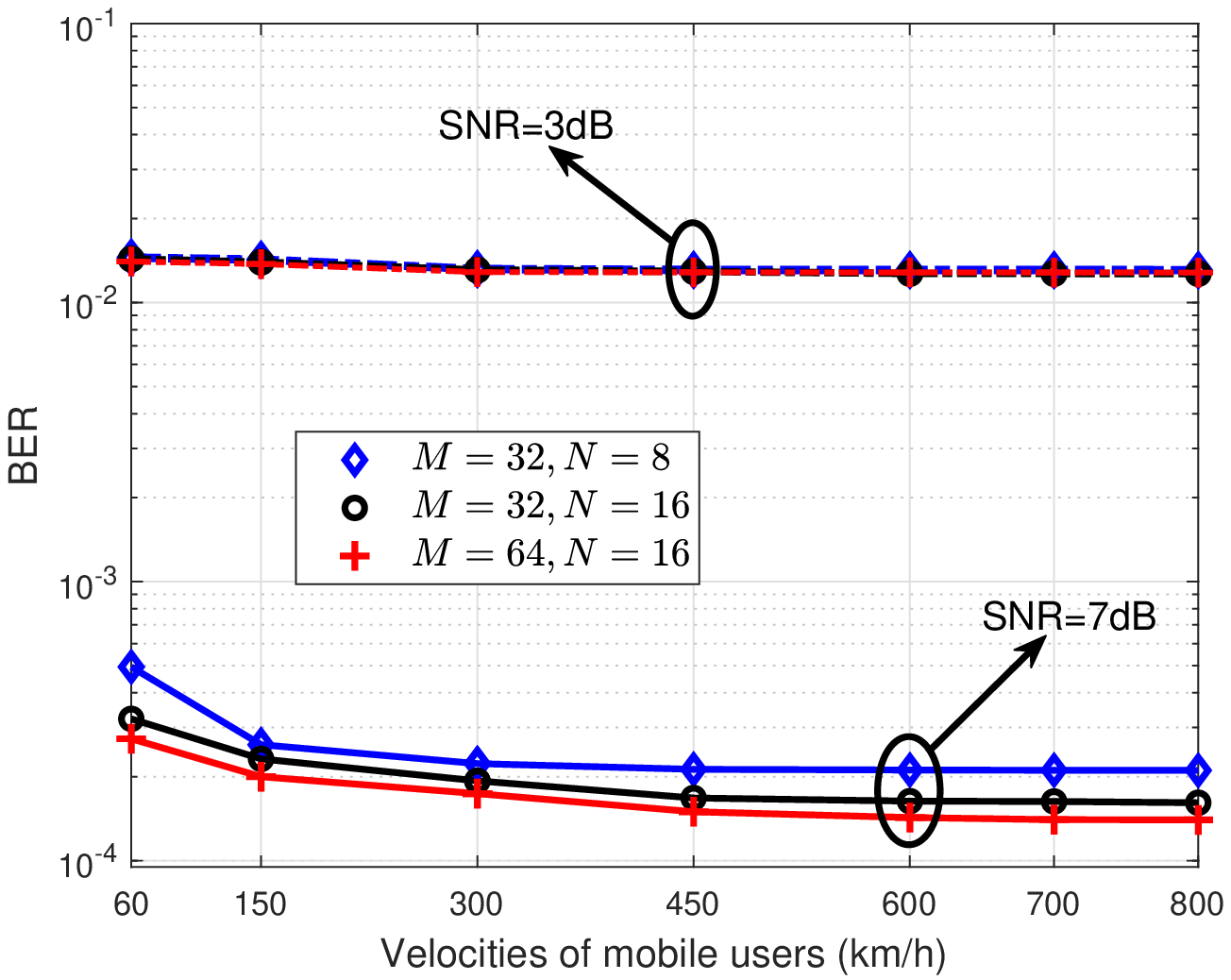}
  \caption{BER performance of Memory AMP for different user velocities and system settings.}
  \label{velocity_MN}
\end{subfigure}
\caption{Performance test of Memory AMP detector for MIMO-OTFS SCMA systems.}
\label{test_MAMP}
\vspace{-1.5em}
\end{figure*}

%\begin{figure}
%  \centering
%  \includegraphics[width=3.6in]{Convergence}
%  \caption{Convergence analysis of Memory AMP with different damping length.}\label{Convergence_damping}
%\end{figure}
We first investigate the convergence and the effects of damping length on Memory AMP receiver performance. Fig. \ref{Convergence_damping} illustrates the bit error rate (BER) performance of Memory AMP versus the number of iterations with different damping length. It can be observed that the BER decreases monotonically and converges within a certain number of iterations. We also notice that no obvious performance improvement after the damping length $L>3$. In the rest of our simulations, we shall use $L=3$ and $\mathcal{T}=6$ for simplicity.

Fig. \ref{antenna_eigenvalue} further tests the effects of the eigenvalue bounds approximation on Memory AMP receiver under different number of BS antennas. It is obvious that the Memory AMP with approximate eigenvalue bounds \cite{9805776} can achieve similar performance as that of exact eigenvalues. This strongly support the effectiveness of such approximation in Memory AMP. We also note that BER performance improves as the number of BS antennas increases due to the additional spatial diversity.
%\begin{figure}
%  \centering
%  \includegraphics[width=3.6in]{Antennas}
%  \caption{BER performance test of the eigenvalue bounds approximation for different number of BS antennas.}\label{antenna_eigenvalue}
%\end{figure}

%\begin{figure}
%  \centering
%  \includegraphics[width=3.6in]{Velocities}
%  \caption{BER performance of Memory AMP for different user velocities and system settings.}\label{velocity_MN}
%\end{figure}
Fig. \ref{velocity_MN} shows the BER performance of the Memory AMP with different user velocities under various settings of $M$ and $N$. As the user velocities increase, the BER performance first improves slightly and then saturates after the velocity beyond $300$ km/h, particularly for high SNR. This is due to the fact that OTFS modulation can resolve a larger number of distinct paths in Doppler domain for higher velocity, leading to performance benefits. We also observe that the BER performance degrades as $M$ and $N$ decrease, especially for the higher SNR. This is attributed to the diversity loss caused by the lower resolution of OTFS delay-Doppler grid.

\begin{figure}
  \centering
  \includegraphics[width=2.9in]{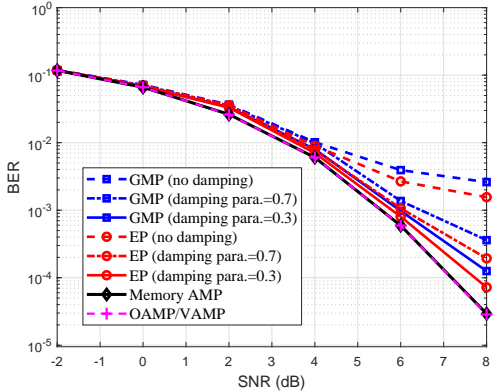}
  \caption{BER performance comparison for different detector algorithms.}\label{detector_compare}
  \vspace{-1.5em}
\end{figure}

Fig. \ref{detector_compare} compares the BER performance of the MIMO-OTFS SCMA systems with different detector algorithms. To highlight the superiority of the proposed Memory AMP algorithm, we also provide the performance of traditional GMP \cite{raviteja2018interference,xiang2021gaussian}, EP \cite{ge2023otfs,9827946} and OAMP/VAMP \cite{ge2021otfs} as benchmarks in Fig. \ref{detector_compare}. The results reveal that
%The results reveal that all the detectors deliver improved performance with higher signal-to-noise ratio (SNR). 
the performance of traditional GMP and EP detectors are very sensitive to the damping parameters. Unfortunately, there is no efficient damping solution for GMP and EP detectors currently.
% which may even cause error floor if no damping is applied. 
However, our developed Memory AMP with closed-form damping solution achieves similar performance to that of OAMP/VAMP even if a low-complexity matched filter is used, and outperforms both the GMP and EP detectors. Based on these analysis, we demonstrate that our Memory AMP detector can yield practical implementation advantage with low complexity and desired performance.

%Finally, the performance of the proposed Memory AMP detector is tested in terms of CSI uncertainty in Fig. X. Here, we capture the CSI errors by adopting the following model:
%\begin{align*}
%{h_{uj,i}} = {{\hat h}_{uj,i}} + \Delta {h_{uj,i}},\;\left\| {\Delta {h_{uj,i}}} \right\| \le {\epsilon _{{h_{uj,i}}}},\\
%{\tau _{uj,i}} = {{\hat \tau }_{uj,i}} + \Delta {\tau _{uj,i}},\;\left\| {\Delta {\tau _{uj,i}}} \right\| \le {\epsilon _{{\tau _{uj,i}}}},\\
%{\nu _{uj,i}} = {{\hat \nu }_{uj,i}} + \Delta {\nu _{uj,i}},\;\left\| {\Delta {\nu _{uj,i}}} \right\| \le {\epsilon _{{\nu _{uj,i}}}},
%\end{align*}
%where ${{\hat h}_{uj,i}}$, ${{\hat \tau }_{uj,i}}$ and ${{\hat \nu }_{uj,i}}$ are the estimated versions of ${h_{uj,i}}$, ${\tau _{uj,i}}$ and ${\nu _{uj,i}}$, respectively. $\Delta {h_{uj,i}}$, $\Delta {\tau _{uj,i}}$ and $\Delta {\nu _{uj,i}}$ denote the corresponding channel estimation errors, whose norms are bounded by the given radius ${\epsilon _{{h_{uj,i}}}} = \epsilon \left\| {{{\hat h}_{uj,i}}} \right\|$, ${\epsilon _{{\tau _{uj,i}}}} = \epsilon \left\| {{{\hat \tau }_{uj,i}}} \right\|$ and ${\epsilon _{{\nu _{uj,i}}}} = \epsilon \left\| {{{\hat \nu }_{uj,i}}} \right\|$, $\forall u,j,i$, respectively. From Fig. X, it is observed that all the detectors experience gradually mild performance loss as the channel uncertainty $\epsilon$ increases. Without sudden and large drop of the receiver performance, our proposed Memory AMP detector exhibits robustness as that of traditional detectors and can handle typical CSI errors. 

\section{Conclusion}\label{VI_conclusion}

In this paper, we developed a low-complexity yet effective customized Memory AMP detector, which is suitable for large-scale MIMO-OTFS SCMA system with inherent channel sparsity. In particular, the large-scale matrix inverse is replaced by a finite terms of matrix Taylor series with low-complexity. We applied the specific orthogonality principle and derived an optimal damping solution in Memory AMP detector for better performance. We also analyzed and compared the computational complexity of the Memory AMP with other benchmark detectors. Our results demonstrated that the proposed Memory AMP detector can yield practical implementation advantage with low complexity and desired performance, and outperforms the existing solutions.

% or
%\appendix % for no appendix heading
%From, we have


%\appendices

%\section{}

\vspace{-0.7em}
\section*{Acknowledgment}
\vspace{-0.4em}
This study is supported under the RIE2020 Industry Alignment Fund—Industry Collaboration Projects (IAF-ICP) Funding Initiative, as well as cash and in-kind contribution from the industry partner(s).

\ifCLASSOPTIONcaptionsoff
  \newpage
\fi



%\bibliographystyle{IEEEtranTCOM}

%\begin{thebibliography}{1}


%\end{thebibliography}

\vspace{-0.9em}
\bibliographystyle{IEEEtran}
%\small
\footnotesize
\bibliography{ref_OTFS_CoMP}

%\begin{biography}[{\includegraphics[width=1in,height=1.25in,clip,keepaspectratio]{mshell}}]{Michael Shell}


%\vfill


\end{document}